\documentclass[10pt,a4paper]{article}
\usepackage{graphicx}
\usepackage{amsmath}
\usepackage{amssymb}
\usepackage{xcolor}

\begin{document}

  \begin{center}
      \Large\textbf{Thin-disk laser pump schemes for large number of passes and
  moderate  pump source quality}\\[4mm]

\normalsize
K. Schuhmann$^{1,2,*}$, T. W. H\"ansch$^3$,
K.~Kirch$^{1,2}$, A.~Knecht$^2$, F.~Kottmann$^{1}$, F.~Nez$^4$,
R.~Pohl$^3$, D.~Taqqu$^{1}$ and A.~Antognini$^{1,2}$\\[3mm]

\textit{
$^1$ Institute for Particle Physics, ETH, 8093 Zurich, Switzerland\\
$^2$ Paul Scherrer Institute, 5232 Villigen-PSI, Switzerland\\
$^3$ Max Planck Institute of Quantum Optics, 85748 Garching, Germany\\
$^4$ Laboratoire Kastler Brossel, UPMC-Sorbonne Universites, CNRS, ENS-PSL Research University,
    College de France, 4 place Jussieu, 75005 Paris, France}\\[3mm]

\textit{skarsten@phys.ethz.ch} %

\end{center}

%% \author{Karsten Schuhmann$^{1,2,*}$,  
%% Theodor W. H\"ansch$^3$,
%% Klaus Kirch$^{1,2}$,
%% Andreas Knecht$^2$,
%% Franz Kottmann$^{1}$,
%% Francois Nez$^4$,
%% Randolf Pohl$^3$,
%% David Taqqu$^{1}$
%% and Aldo Antognini$^{1,2}$

%% }

%% \address{$^1$ Institute for Particle Physics, ETH, 8093 Zurich, Switzerland\\
%% $^2$ Paul Scherrer Institute, 5232 Villigen-PSI, Switzerland\\
%% $^3$ Max Planck Institute of Quantum Optics, 85748 Garching, Germany\\
%% $^4$ Laboratoire Kastler Brossel, UPMC-Sorbonne Universites, CNRS, ENS-PSL Research University,
%%    College de France, 4 place Jussieu, case 74 75005 Paris, France}

%% \email{$^*$ skarsten@phys.ethz.ch} %% email address is required

%% % \homepage{http:...} %% author's URL, if desired

%% %%%%%%%%%%%%%%%%%%% abstract and OCIS codes %%%%%%%%%%%%%%%%
%% %% [use \begin{abstract*}...\end{abstract*} if exempt from copyright]

\begin{abstract}
Novel thin-disk laser pump layouts are proposed yielding an increased
number of passes for a given pump module size and pump source quality.
These novel layouts result from a general scheme which bases on 
merging two simpler pump optics arrangements.
Some peculiar examples can be realized by adapting standard
commercially available pump optics simply by introducing an additional
mirror-pair.
More pump passes yield better efficiency, opening the way for usage of
active materials with low absorption.
In a standard multi-pass pump design, scaling of the number of beam
passes brings about an increase of the overall size of the optical
arrangement or an increase of the pump source quality requirements.
Such increases are minimized in our scheme, making them eligible for
industrial applications.
\end{abstract}

%% \ocis{ (140.3615)   Lasers, ytterbium,
%%        (140.4480)   Optical amplifiers, 
%% %       (140.6810)   Thermal effects,   
%%        (140.3325)   Laser coupling, 
%%        (140.3580)   Lasers, solid-state,
%%        (140.5560)   Pumping,
%%        (080.3620)   Lens system design,
%%        (080.4035)   Mirror system design,

%%  }
 % REPLACE WITH CORRECT OCIS CODES FOR YOUR ARTICLE, MINIMUM OF TWO; Avoid using the OCIS codes for “General” or “General science” whenever possible.
 %For a complete list of OCIS codes, visit: http://www.opticsinfobase.org/submit/ocis/

\section{Introduction}

\begin{figure}[h!]
\centering\includegraphics[width=\textwidth]{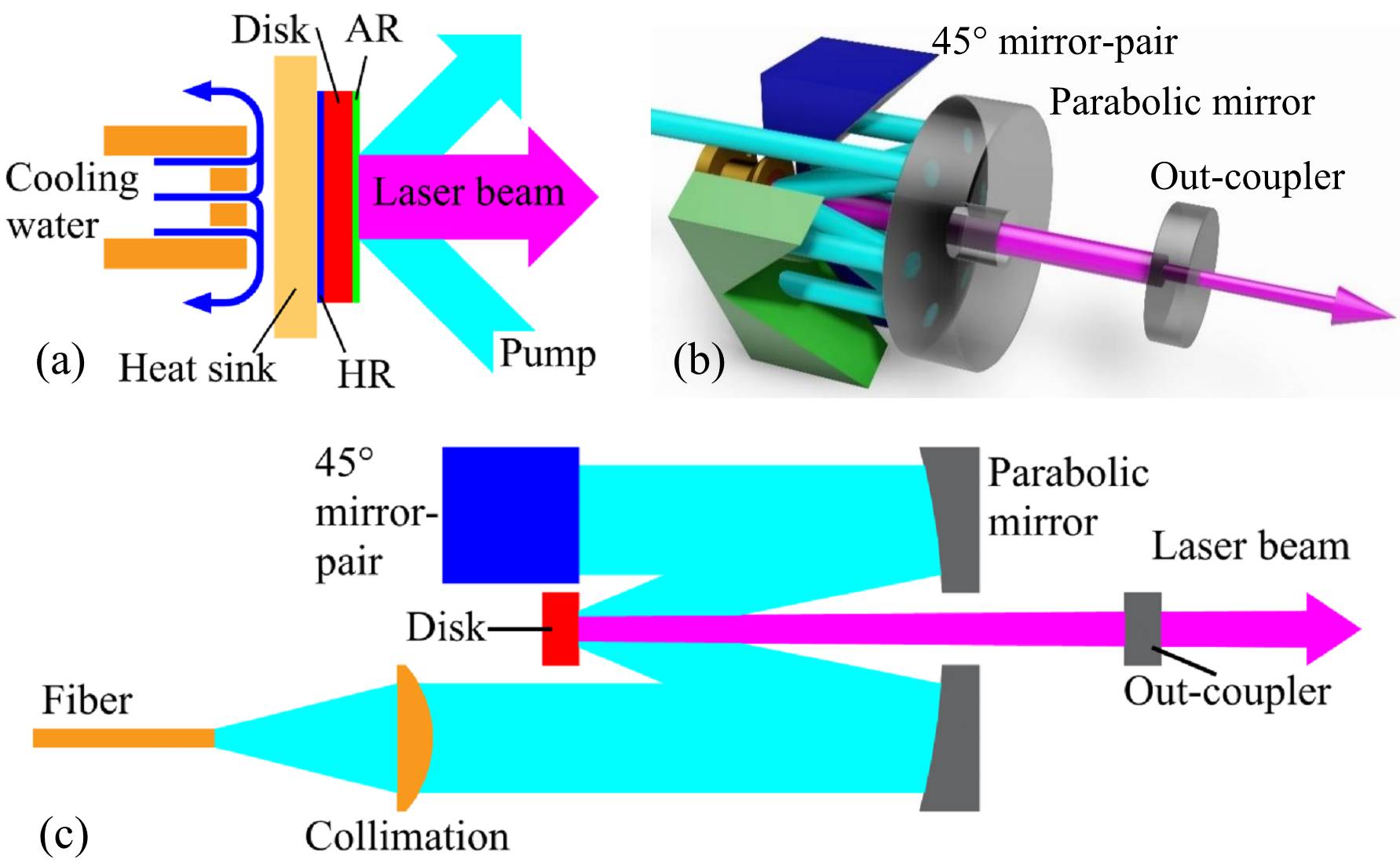}
\caption{\label{fig:1} \it (Color online) Working principle of a thin-disk
  laser and thin-disk laser pump arrangement. (a) Scheme of the
  thin-disk active medium mounted on a water cooled heat sink. Lasing
  and cooling occurs along the disk axis. (b) 3D schematic of the pump
  optics: heat sink (gold), parabolic mirror (gray), and prisms acting as
  mirror-pairs (green and blue). (c) Pump light multi-pass
  arrangement. The light from an homogenizer is imaged onto the disk via
  a parabolic mirror. The multi-pass is realized via the disk, the
  parabolic mirror and the 45$^\circ$ mirror-pairs.  In cyan is
  given the pump beam propagation in the multi-pass,   in magenta the laser
  beam, HR and  AR stand for high-reflector and  anti-reflex layer, respectively. }
\end{figure}

The thin-disk laser~\cite{Giesen1994, Brauch1995, Stewen2000} is a diode-pumped
solid-state laser with high power and high pulse energy capabilities,
high efficiency and excellent beam quality.
The thin-disk laser active medium, depicted in Fig.~\ref{fig:1}~(a),
is a thin disk with typical thickness of 100-500~$\mu$m and diameter
up to a few cm.
Lasing and cooling occurs along the disk axis, while pumping is ensued
in a quasi-end-pumped configuration.
The rear side of the disk is acting as a high-reflective mirror (HR)
for pump and laser wavelengths, and it is thermally coupled to 
a heat sink.
Heat removal from the disk is efficient because of the large cooled
surface to active volume ratio.
Since the heat flux occurs along the laser axis, the thermally induced
lens effects are minimized resulting in small phase-front distortions
also for beams of large
diameter~\cite{Antognini2009, Speiser2007, Zhu2014, Perchermeier2013}.
This cooling scheme allows thus for power and energy
scaling~\cite{Speiser2009, Giesen2007, Piehler2012,
  Mende2009, Fattahi2014,  Saraceno2015} simply by increasing the diameter of laser and pump
spots, eventually limited by amplified spontaneous emission 
effects~\cite{Antognini2009, Speiser2009a, Furuse2013, Peterson2011}.
Moreover, the efficient cooling allows pumping in the kW regime and
the usage of quasi three-level-system materials having low quantum
defect and high gain.
To date, the paradigmatic material of choice, especially in industrial
applications, is ytterbium-doped yttrium aluminium garnet
(Yb:YAG)~\cite{TRUMPF_www, Gottwald2012, Larionov2014} but recently,
researchers have concentrated on finding new materials with larger
thermal conductivity for higher output power, and with larger
bandwidth for ultrashort pulse generation or tunable
lasers~\cite{Giesen2007, Peters2007, Peters2009, Siebold2012, 
   Saraceno2013, Richaud2011, Dannecker2014}.

The small thickness of the disk guarantees excellent cooling and power
scaling.
Yet, in a single pass, only a small fraction of pump light is absorbed
in the thin active medium.
The light not absorbed in the disk in the first pass is reflected by
the HR coating onto a second pass  in the active medium.
Even so, the absorption in the resulting double pass is too small.
This shortcoming can be compensated using a multi-pass scheme for the
pump light, that is, by redirecting the (not absorbed) pump beam into
the disk several times~\cite{Erhard2000}.

A well defined pump region with sharp boundaries is fundamental,
especially for three-level-system materials due to the high lasing
threshold.
To generate a pump profile which minimizes radial tails, on one hand
the pump optics has to redirect the various passes at ``exactly'' the
same position at the disk, and, on the other hand, each individual
pump pass spot has to have sharp boundaries.
Relay 4f telecentric imaging with unitary magnification is commonly
used for this purpose~\cite{Erhard2000} as its propagation matrix is
exactly the negative unity matrix.

The scheme ordinarily used for disk laser pumping is first to
homogenize a high-power diode laser by coupling it into either a
multi-mode fiber or a rod homogenizer.
The output face of this optical element is then imaged with suitable
magnification onto the disk.
The multi-pass propagation of the pump beam is realized using
telecentric 4f relay imaging with unitary magnification, that is, by
imaging the pump spot at the disk position from pass to pass without
changing its size.
The imaging properties of the 4f imaging scheme ``per definition'' guarantees
that the beam spot profile and its divergence are exactly reproduced
from pass to pass (neglecting phase-front distortions occurring at the
disk).

Another practical advantage is given by the fact that 4f imaging can
be realized with only a few optical elements also for a large number of
passes.
Commonly it is realized~\cite{Erhard2000, Contag2000_patent1,
  Erhard2000_patent2, Erhard2001_patent3,Erhard2001b_patent4} using a parabolic mirror and a deflecting mirror
system as shown in Fig.~\ref{fig:1} (b) and (c).
After collimation, the pump light from the homogenizer enters the pump optics
and is being directed  to the disk by the parabolic mirror.
The light not absorbed in the disk in the first double-pass
is reflected back to the parabolic mirror.
At the parabolic mirror, the pump beam is collimated and sent to the
mirror system (prisms) which after a double 90$^\circ$ reflection
redirects the light back onto the parabolic mirror, but at a different
position.
From there, the light proceeds to the disk for the second time.
Iterating this scheme several times gives rise to a multi-pass pump
pattern having several passes through the active medium.

The typical commercially available thin-disk pump
modules~\cite{DG_www1, DG_www2, IFSW_www1} provide 24 or 48
passes\footnote{Throughout this paper, the number of passes $N$
is defined  to be twice the number of reflections at the disk.}.
\textcolor{black}{
A larger number of pump passes enables a reduction of the thin-disk
thickness and doping concentration while keeping the same pump light absorption.
The smaller heat resistance that results from reducing the thin-disk
thickness brings about lower average temperature of the  active
medium which is advantageous for the three-level-systems.
Moreover also thermal lens effects (spherical and aspherical
components) of the thin-disk, are reduced leading to improved efficiency and beam
quality.
Similarly, a reduced doping ensues higher active medium thermal conductance 
due to decreased  scattering of the active medium phonons on the doping ions
yielding a more efficient cooling.
The increase of pump passes also allows for the use of active
medium materials with smaller absorption cross sections,
and smaller absorption bandwidths.
}

However, such an increase of pass number has to be attained without
increasing substantially the requirements for the pump beam quality
expressed by the beam parameters product $P$ which is defined as the
product of the pump beam's divergence angle (half-angle in the
far-field) and beam waist (radius of the beam at its narrowest point).

In this paper, possible realizations of multi-pass pump layouts suited
for thin-disk lasers are presented.
We restrict the discussion to 4f imaging schemes.
In Sec.~\ref{sec:lens} the 4f imaging and the multi-pass concepts are
introduced with the help of an exemplary simple pump layout.
Section~\ref{sec:state-of-the-art} describes the state-of-the-art of
the standard commercially available thin-disk pump optics and its
related pump beam propagation.
The novel layouts we are proposing in this work, result
from combining two multi-pass concepts: the first is based on the
standard pump module design described in
Sec.~\ref{sec:state-of-the-art}, the second is given in
Sec.~\ref{sec:circles}.
The merging of these two concepts which is described in
Sec.~\ref{sec:merging} enables the realization of multi-pass systems
with large number of passes while only minimally increasing the
complexity of the multi-pass system, the size of the optics, and the
requirement for the pump source quality, making these schemes apt for
industrial applications.
A particular practical realization based on hexagonal mirror-pairs and
hexagonal ordered fiber-coupled diodes is presented in
Sec.~\ref{sec:hexagonal}, followed by concluding remarks.

%%%%%%%%%%%%%%%%%%%%%%%%%%%%%%%%%%%%%%%%%%%%%%%%%%%%%%%%%%%%%%%%%%%%%%%%%%%%%%%%
\section{A simple example of a multi-pass scheme based on 4f relay imaging}
\label{sec:lens}
\begin{figure}[h]
\centering\includegraphics[width=0.9\textwidth]{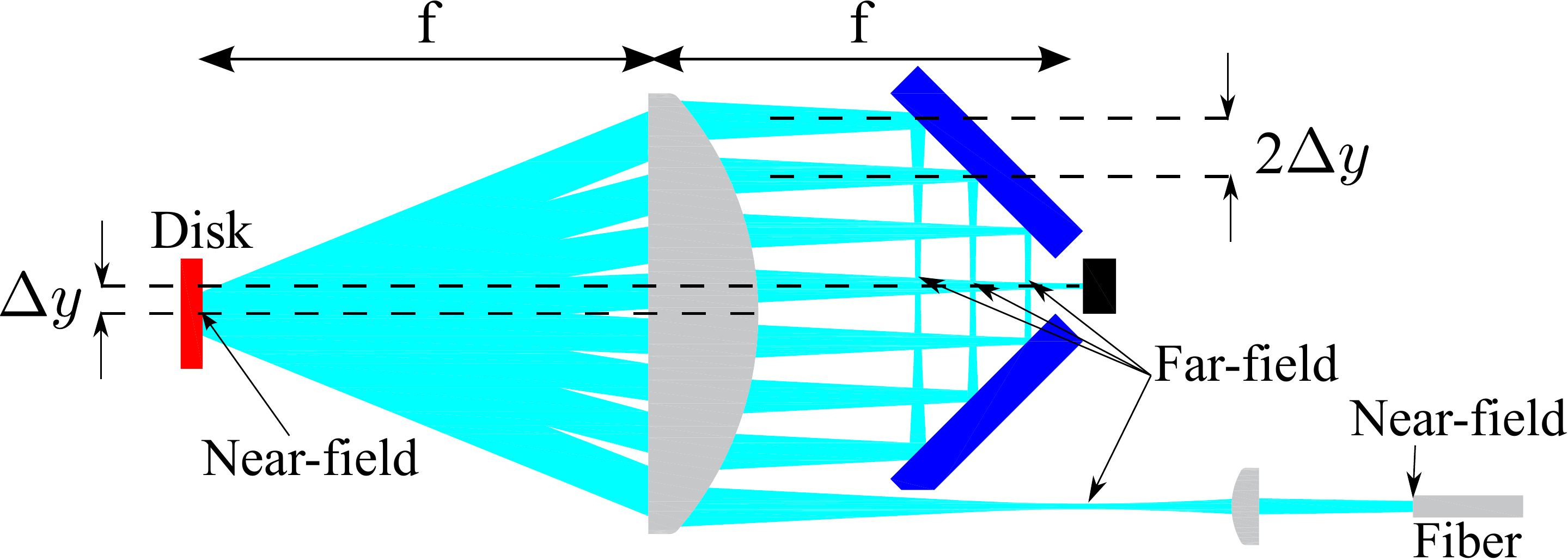}
\caption{\label{fig:2} \it (Color online) Schematic of a 4f multi-pass optical system
  realized by a 45$^\circ$ mirror-pair (blue), a lens (gray)
  and a disk (red).  The mirror-pair axis  is shifted by $\Delta y$
  relative to disk-lens axis. The pump beam multi-pass is given in
  cyan.  The number of passes can be doubled by introducing a
  back-reflector (black). }
\end{figure}
A simple way to realize a multi-pass scheme based on 4f imaging is
shown in Fig.~\ref{fig:2}~\cite{Giesen-2012_patent5, DG_www3}.
The setup consists of a thin-disk, a lens with focal length $f$ and a
pair of mirrors at 45~$^\circ$ angle relative to the optical axis.
The spacings between  disk and lens, and between lens
and the intersect of the mirror-pair are $f$.
Hence, the propagation from disk to mirror-pair can be seen as imaging
from the disk image plane to the mirror-pair Fourier plane, and the
full round-trip from disk to disk as a 4f relay imaging.
In other words, the near-field laser profile at the disk is imaged by
the lens into the far-field at the mirror-pair center and vice versa.

If the intercept (center) of the mirror-pair was located  on the disk-lens axis,
the light would propagate in this optical system in a closed loop as
in a resonator.
However, by displacing the mirror-pair center by $\Delta y$ as shown
in the figure, a multi-pass configuration can be realized.
The resulting far-field spacing of the various beam passes  is $2\Delta y$.
The shift of the mirror-pair center relative to the common disk-lens
axis breaks the symmetry which is necessary for the realization of a
multi-pass scheme.
The size of $\Delta y$  controls the number of achievable passes given
constraints from the laser beam size, restrictions related with
astigmatisms and the size of the various optical elements.

A common feature of multi-pass propagation is that the number of
passes can be doubled by placing a back-reflector at its end causing
the beam to retrace itself traveling in the opposite direction.
For pump beams, the residual beam not absorbed in the disk after the
back-and-forth propagation is consequently sent back to the
homogenizer of the pump diodes.
%% %

\section{State of the art of the commercial multi-pass pump systems}
\label{sec:state-of-the-art}

\begin{figure}[htbp]
\centering
    \setlength{\unitlength}{1.0mm}
    \begin{picture}(120,35)(0,0)
     %\put(0,0){\framebox(120,35)}
     \put(-3,-1){\includegraphics[width=0.33\textwidth]{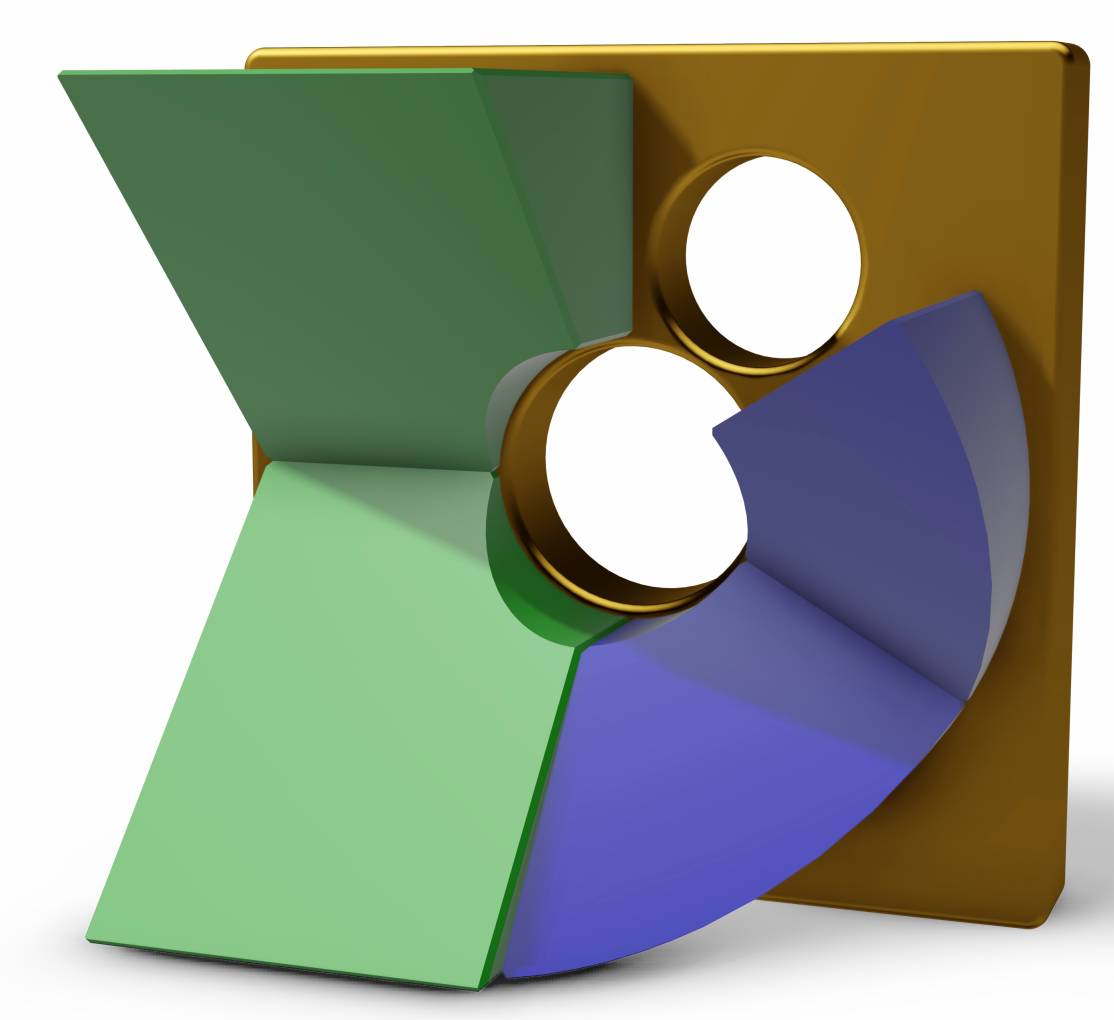}}
     \put(40,0){\includegraphics[width=0.36\textwidth]{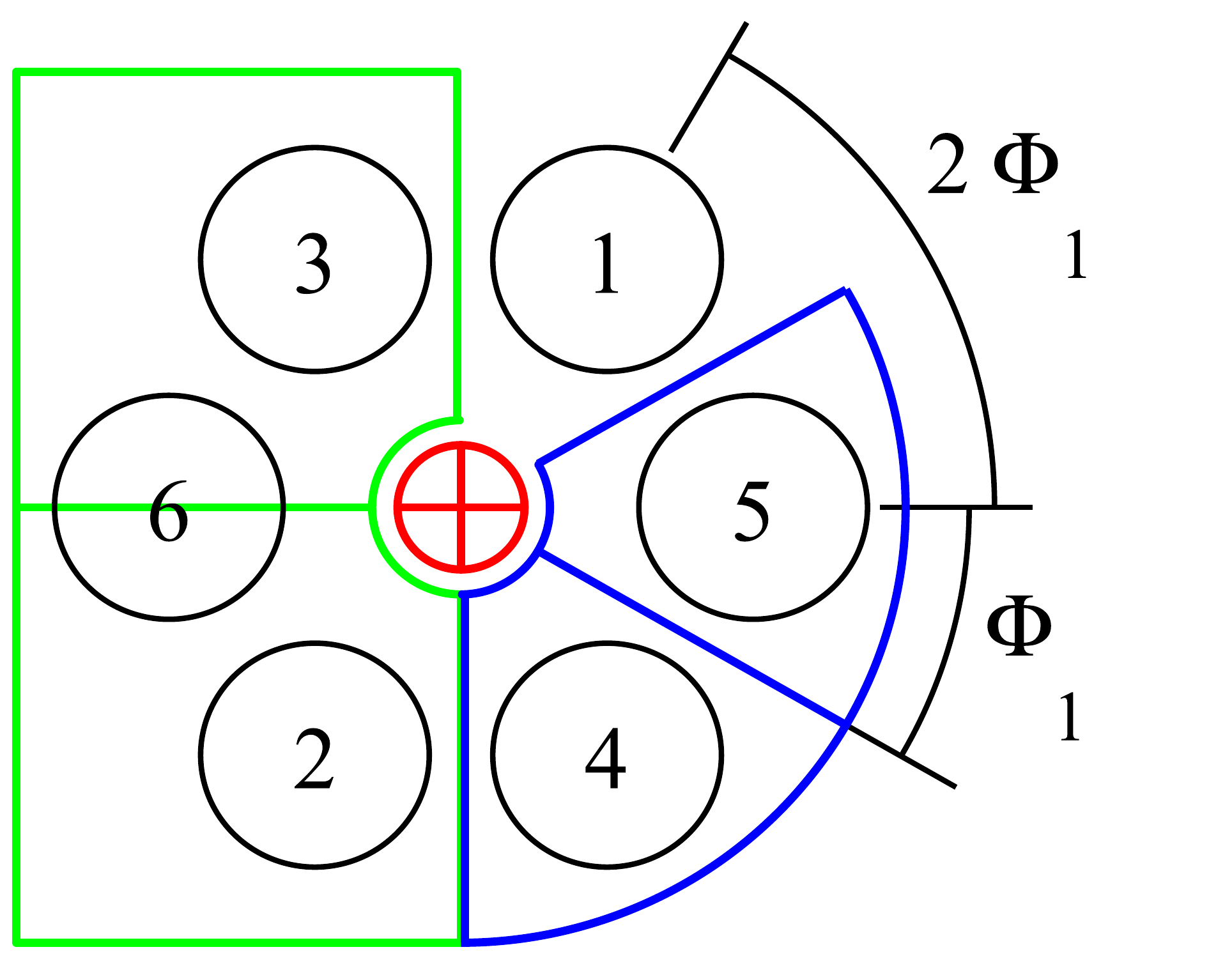}}
     \put(83,-2){\includegraphics[width=0.37\textwidth]{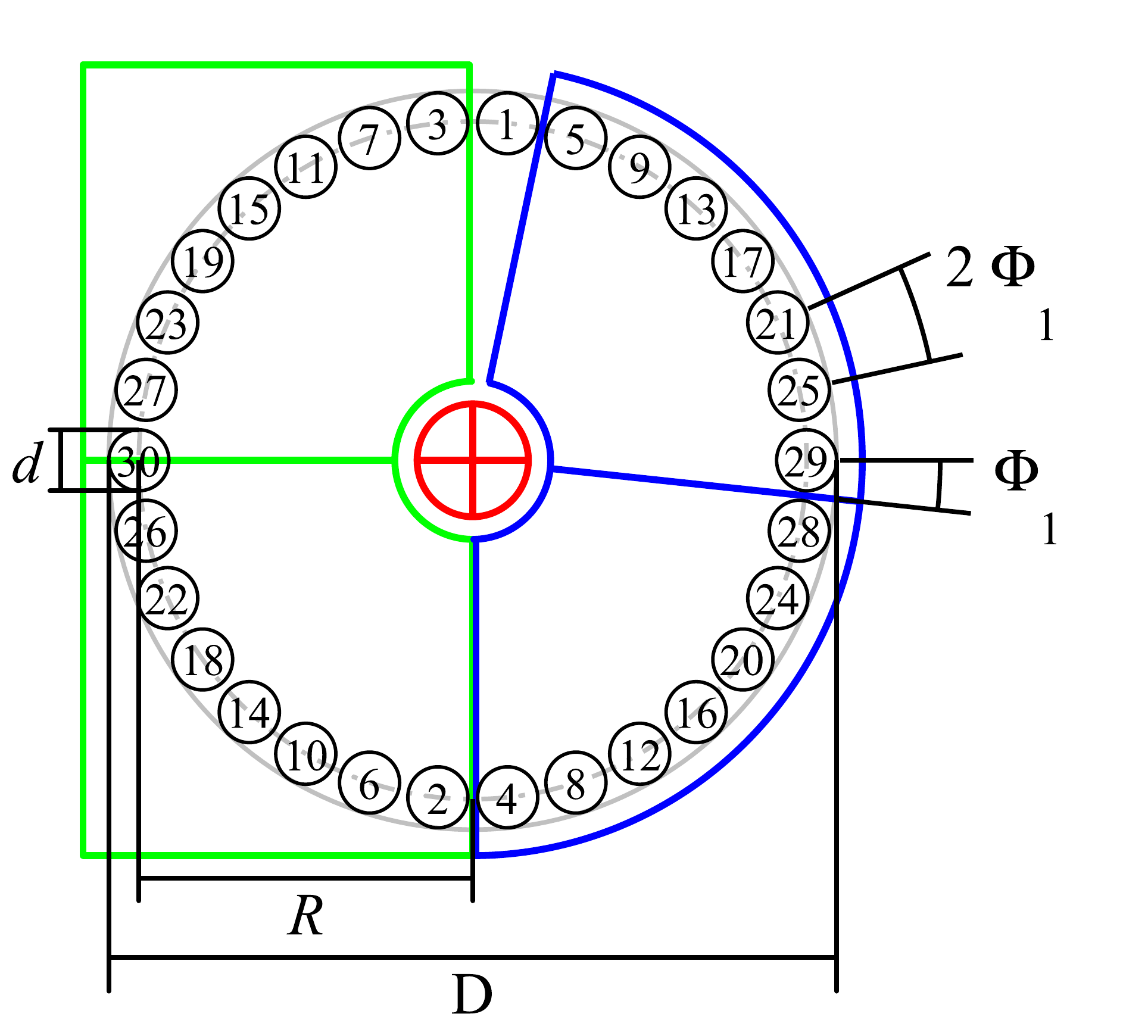}}
    \put(3,2){\large (a)}
    \put(41,2){\large (b)}
    \put(80,2){\large (c)}
    %\put(70,5){\large (d)}
\end{picture}
\caption{\label{fig:3} \it Schematics of the mirror-pairs and multi-pass
  beam routing at the parabolic mirror and mirror-pairs plane.  (a)
  3D arrangement of the mirror- (prisms-) pairs commercially
  available. (b) Mirror-pairs contours and beam
  routing (which follows the given numbering) for 6
  reflections (12 passes) at the disk. The disk position is given in
  red.  The second mirror-pair (blue) is rotated by an angle
  $\phi_1=30\,^\circ$ relative to the first mirror-pair (green).  At
  position 6 the beam is back-reflected doubling the number of passes.
  (c) Schematic for 30 reflections (60 passes) at the disk achieved by
  decreasing the angle $\phi_1$.  }
\end{figure}
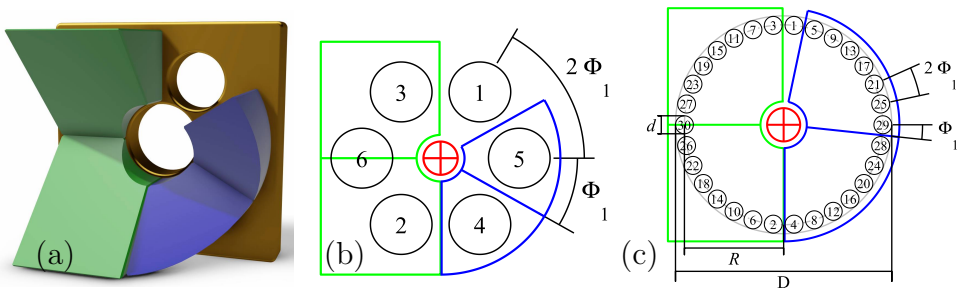

A typical pump optics system for thin-disk lasers originally proposed
in~\cite{Erhard2000} and now available at TRUMPF, IFSW and
Dausinger+Giesen is given in Fig.~\ref{fig:1}.
It consists of a disk, a large parabolic mirror with focal length $f$
and two mirror-pairs (HR coated prism-pairs).
Similar to the previous example, the distances between disk and parabolic
mirror, and between parabolic mirror and prisms  intersects
are $f$.
For the realization of a multi-pass propagation the symmetry of the
optical layout has to be broken.
Instead of shifting as previously the mirror-pair off-axis, in this
case, the symmetry is broken by rotating one prism-pair by an angle
$\phi_1$ relative to the other pair as depicted in Fig.~\ref{fig:3}.

To understand the beam routing in the multi-pass pump optics consider
Fig.~\ref{fig:3} (b).
The position of the pump beam at the mirror-pairs plane (and at the
parabolic mirror as the collimated pump beams are parallel) is
indicated by the numbering.
The number 1 represents the position of the collimated in-coupled
beam.
After a reflection at the parabolic mirror the beam travels to the
disk (red central circle).
The remaining pump light reflected from the disk subsequently reaches the position 2
at the first mirror-pair after a second collimation at the parabolic
mirror.
From there the pump beam is redirected within the same mirror-pair to reach
position 3.
Thereafter, it is sent towards the parabolic mirror, disk,
parabolic mirror to reach the position 4 on the second mirror-pair.
Via a path within this second mirror-pair the beam reaches position 5
where it is redirected again towards the disk (third reflection at the disk)
and from there eventually to position 6.
At position 6 the beam is back-reflected  giving rise to a
propagation in opposite direction which results in a doubling of the
number of pump passes on the disk.

The relative rotation of the mirror-pair intercepts by an angle
$\phi_1$ gives rise to a rotational symmetric beam spot pattern with
spot-to-spot angle of $2\phi_1$.
In this design, all far-field beams have the same radial distance from
the optical axis resulting in a central region free of pump beam spots
(near-field is used for beams at the disk, far-field for beams at the
mirror-pairs plane, as defined in Fig.~\ref{fig:2}).
Therefore, the mechanics holding the mirror-pair can have a central
aperture as shown in Fig.~\ref{fig:3} (a) in which the thin-disk can be
placed.
Similarly, the parabolic mirror can have a central aperture as shown
in Fig.~\ref{fig:1} (b) and (c) for laser beam access. 
In such a way, the radial symmetric beam pattern allows the usage of a
parabolic mirror instead of a lens as in example of Fig.~\ref{fig:2}, which
yields a folding of the beam propagation so that geometrically the
mirror-pairs plane coincides with the disk plane.
%% %
Thus, the utilization of a parabolic mirror reduces the size of the
multi-pass pump optics, leads to smaller absorption losses and offers
the possibility for elegant laser beam coupling (see off-axis aperture
in Fig.~\ref{fig:3} (a)).

The most natural way to increase the number of pump passes would be to
reduce the angle $\phi_1$ as shown in Fig.~\ref{fig:3} (c).
The number of passes $N$ is given by $N=2\times 180/\phi_1$, where
$\phi_1$ is expressed in degrees.
The factor of 2 originates from the fact that each reflection
corresponds to two passes in the active material.
Furthermore, this equation assumes that the pump beam propagates back
and forth in the multi-pass segment due to the back-reflector (placed
e.g., at position 6 in Fig.~\ref{fig:3} (b), or at position 30 in
Fig.~\ref{fig:3} (c)).

However, an increase of passes by decreasing the angle $\phi_1$,
(assuming the same parabolic mirror size and focal length $f$) can be
realized only by decreasing the far-field beam spot size.
Overlapping of the far-field spots (spots at the mirror-pairs plane) is not
acceptable because it implies aperture losses at the beam in-coupling.
For large $N$ the maximal diameter $d$ of a single far-field beam
spot, as can be seen in Fig.~\ref{fig:3}, is given by $d=2\pi R/N$,
\textcolor{black}{where $R$ is the radial distance of
    the pump beams from the optical axis at the parabolic mirror
    position.}
%where $R$ is the radial distance from to optical axis of the pump
%spots in the far-field.
%
This implies $1/N$ scaling of the pump beam parameters product and a poor
usage of the mirror-pairs and parabolic mirror surfaces.
On the contrary, if the total surface of the parabolic mirror was
used, the diameter of the far-field spots would shrink
approximately  only as $1/\sqrt{N}$.

A scheme which makes use of a much larger fraction of the parabolic
mirror surface is presented in the Sec.~\ref{sec:merging}.
This scheme allows thus for scaling of the number of passes while only
moderately increasing the demands on the quality of the pump source.
Before describing this scheme, for didactic reasons in
Sec.~\ref{sec:circles} a multi-pass layout is presented which
forms one of the building blocks of the final schemes.

\section{Two mirror pairs whose intersects meet off-axis}
\label{sec:circles}

\begin{figure}[h!]
\centering
\centering
    \setlength{\unitlength}{1.0mm}
    \begin{picture}(120,63)(0,0)
     %\put(0,0){\framebox(120,63)}
     \put(0,-2){\includegraphics[width=0.6\textwidth]{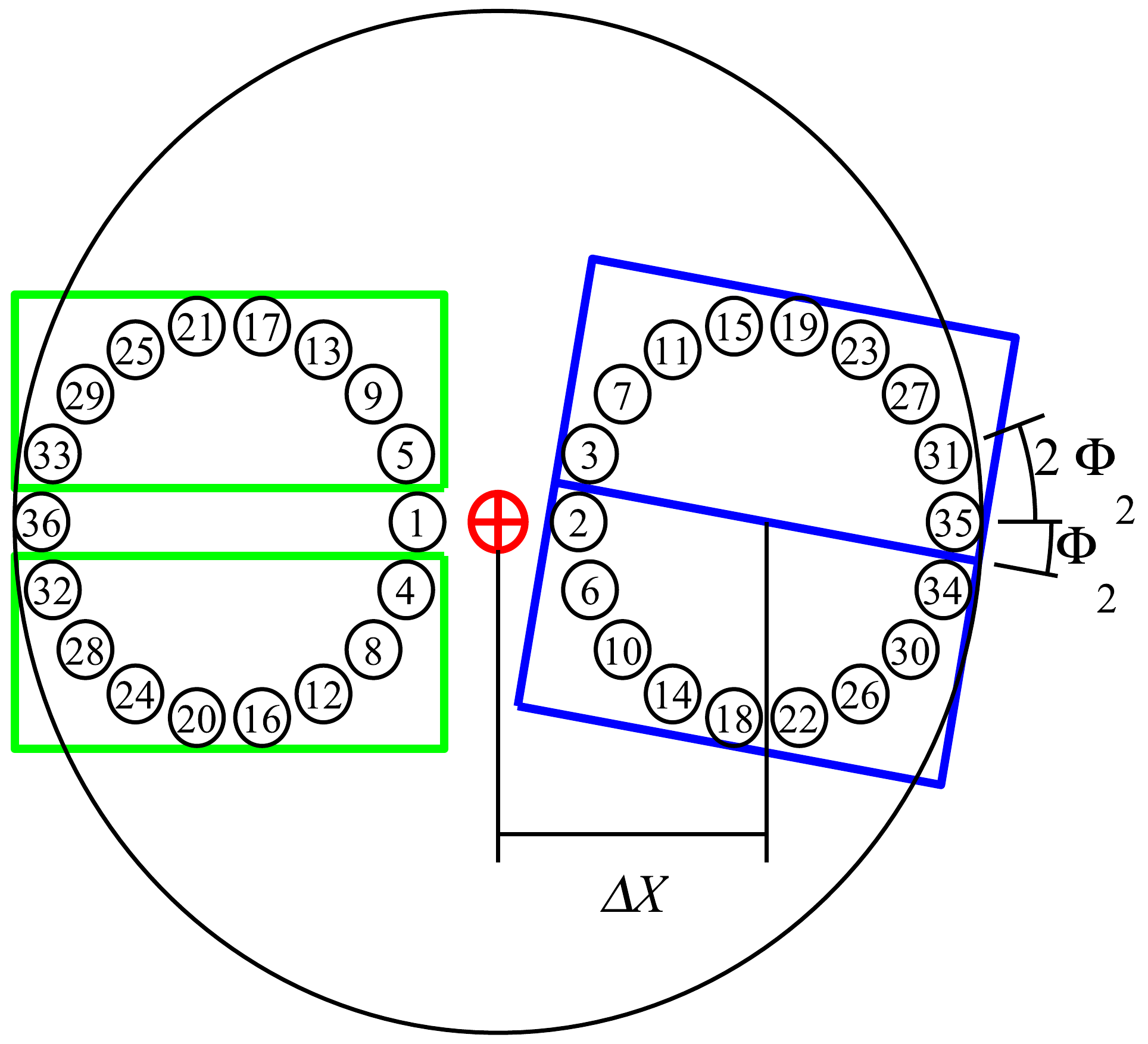}}
     \put(73,-10){\includegraphics[width=0.39\textwidth]{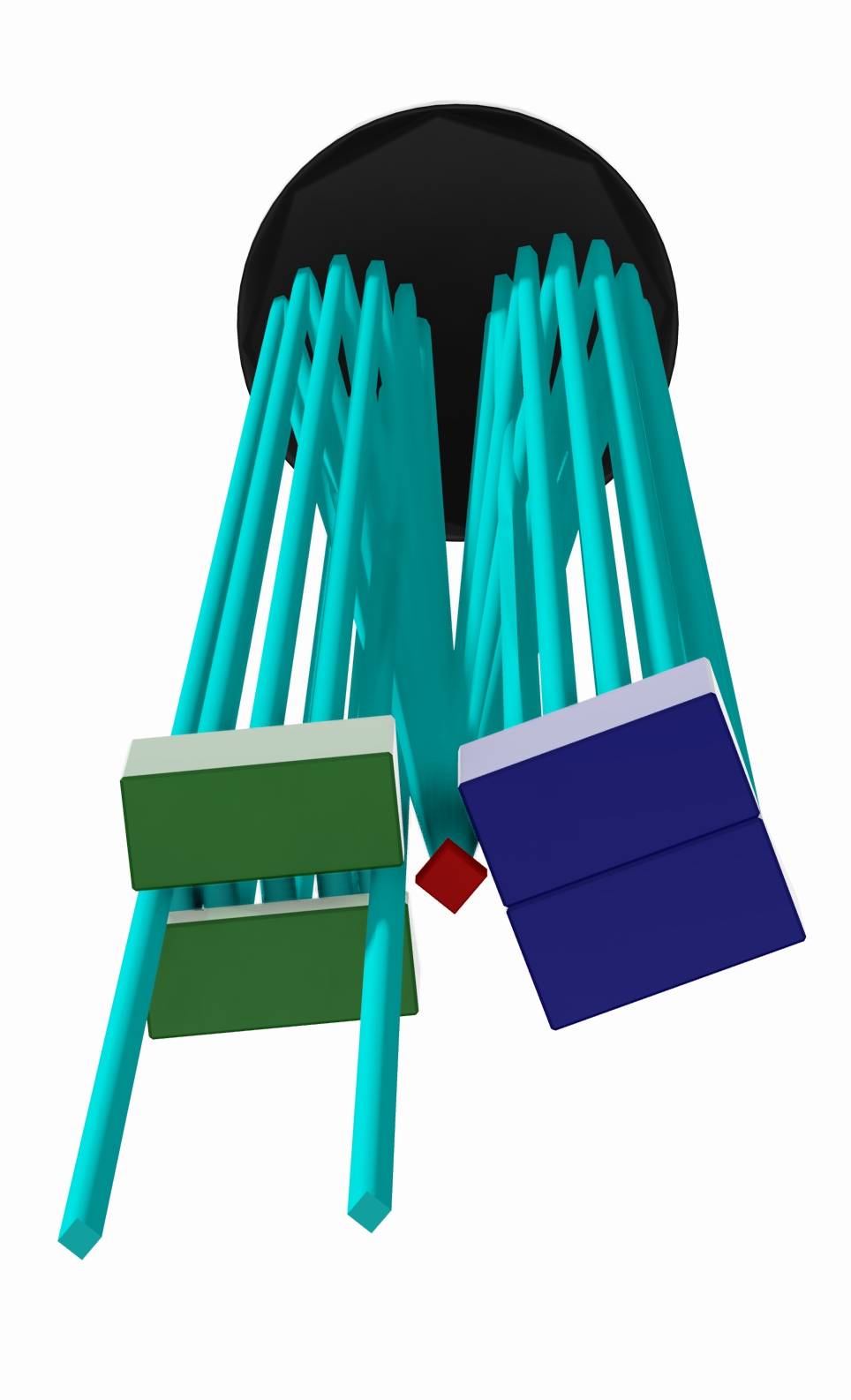}}
    \put(3,2){\large (a)}
    \put(69,2){\large (b)}
    %\put(3,4){\large (c)}
    %\put(70,5){\large (d)}
\end{picture}
\caption{\label{fig:4} \it (a) Multi-pass configuration resulting from two
  mirror-pairs (prism-pairs) whose intersects do not cross at the
  optical axis but at a point with $\Delta X$ offset. (b)
  Corresponding 3D drawings.  Mirror-pairs are given in blue and
  green, the disk in red and the parabolic mirror in gray.}
\end{figure}

The generalized pump schemes we propose and that are detailed in next
section are resting on two building blocks.
The first one is given by the standard pump optics described in
Sec.~\ref{sec:state-of-the-art}.
The second one, described in this section, is a novel design realized
also only with a disk, a parabolic mirror and two mirror-pairs as
shown in Fig.~\ref{fig:4}.
Differently to the layout of Fig.~\ref{fig:3}, the second mirror-pair
is rotated relative to the first one by an angle $\phi_2$ not around
the disk axis but around an axis shifted by an offset $\Delta X$ as
shown in the Fig.~\ref{fig:4} (a).
The resulting distribution of the far-field spots is located on two
separated circles.
Similar to previous configurations, the number of passes is dictated by
the angle of rotation, whereas the radius of these circles is
given by the distance of the in-coupled beam to the center of the
first mirror-pair depicted in green.

The beam is traveling between mirror-pairs, parabolic mirror and disk.
The beam routing starts from the aperture (slit between the green
mirror pair) at position 1 of the first mirror
pair, and reaches position 2 at the second mirror-pair (blue) after a
reflection at the disk and two reflections at the parabolic mirror.
From position 2, the beam is redirected within the second mirror-pair
to position 3.
From position 3, it travels back to the first mirror-pair via a reflection at the disk
and two reflections at the parabolic mirror, to reach position 4.
Within the first mirror-pair the beam travels from position 4 to
5, and from there it is sent again towards the disk.
Iterating this scheme, several passes at the disk are realized,  while
the beam  position at the mirror-pairs travels the given numbering.

For large number of passes, this multi-pass scheme requires
large pump optics and shows a poor utilization of the parabolic mirror
surface.
Therefore, when used as shown in Fig.~\ref{fig:4}, it is not suited
for the realization of a compact pump optics.
However, when combined with other schemes as detailed in next section,
layouts with more efficient usage of the parabolic surface can be
realized.
Still, this scheme as such (as in Fig.~\ref{fig:4}), can be used to design multi-pass laser
amplifiers with large number of passes, because of the typically
better beam parameter products of the laser beams and the less
stringent size limitations.

\section{Novel design with many more passes }
\label{sec:merging}

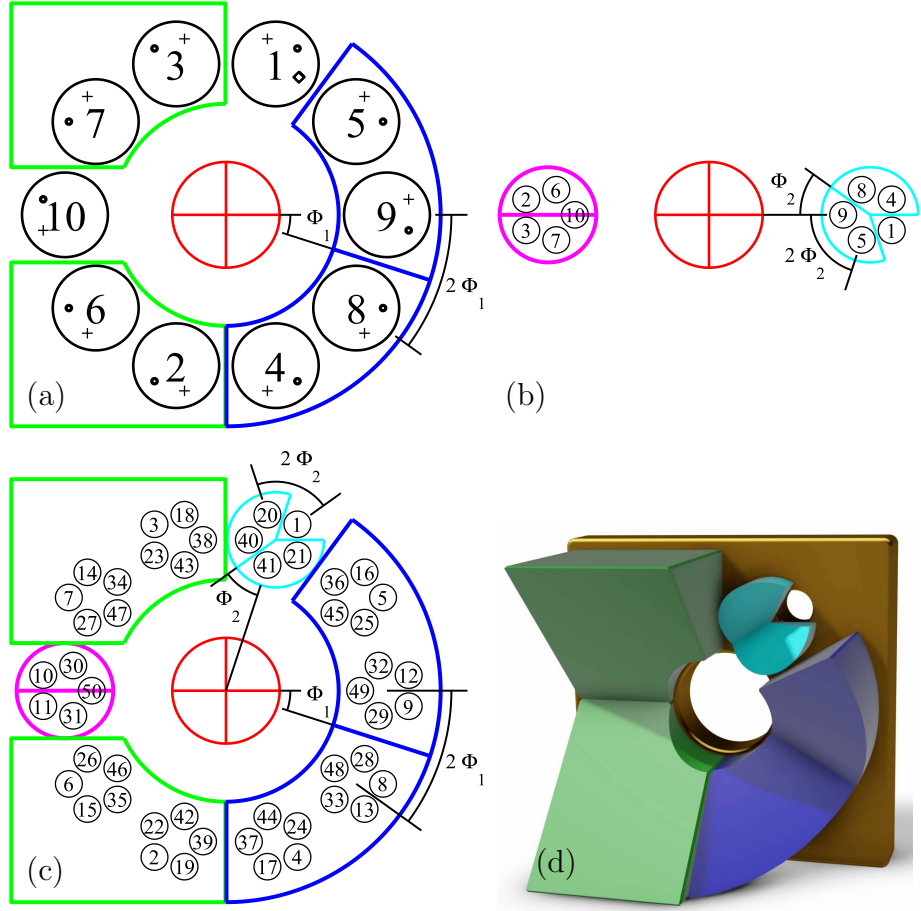
\begin{figure}[htbp]
\centering
    \setlength{\unitlength}{1.0mm}
    \begin{picture}(120,120)(0,0)
     %\put(0,0){\framebox(120,120)}
     \put(0,0){\includegraphics[width=\textwidth]{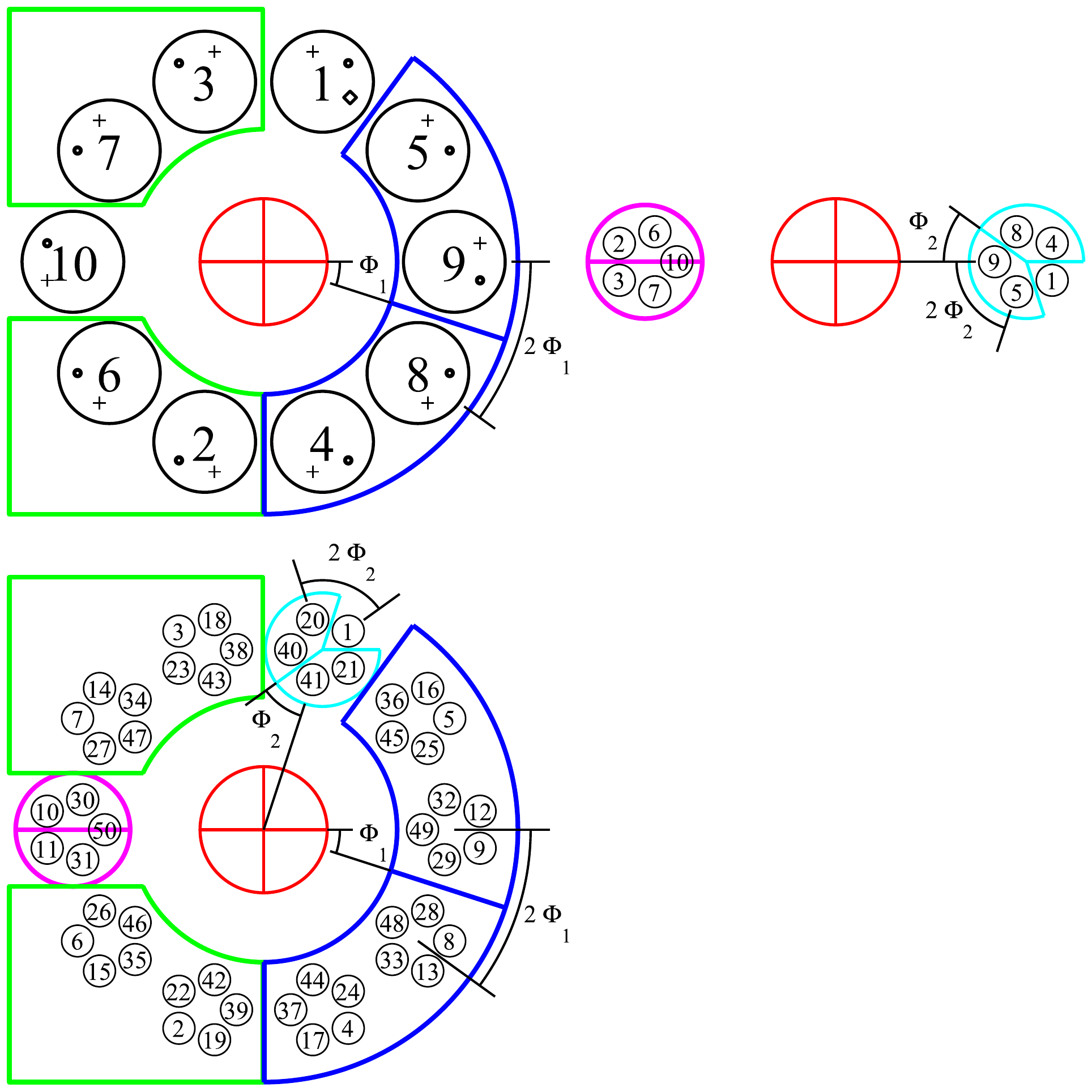}}
     \put(65,0){\includegraphics[width=0.45\textwidth]{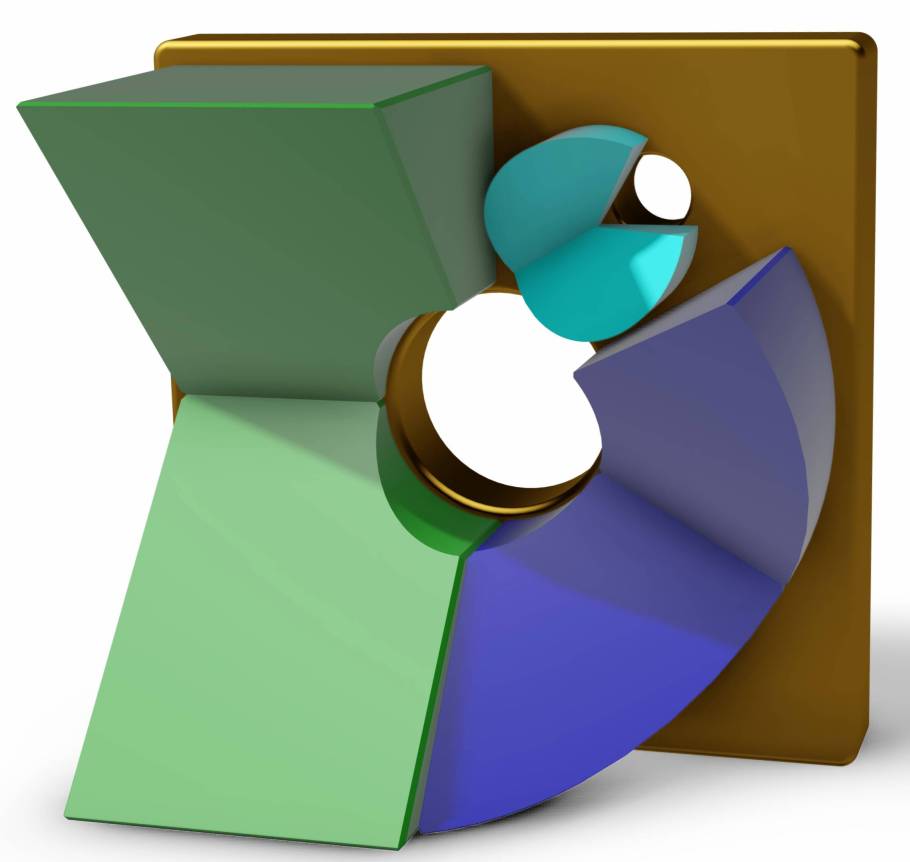}}
    \put(3,67){\large (a)}
    \put(66,67){\large (b)}
    \put(3,4){\large (c)}
    \put(70,5){\large (d)}
\end{picture}
\caption{\label{fig:5} \it Principle of the merging of the two multi-pass
  schemes of Fig.~\ref{fig:3} and Fig.~\ref{fig:4}.  (a) Similar
  mirror-pair scheme as in Fig.~\ref{fig:3} for 20 passes. The slightly
  different position of the in-coupled light in region 1 is producing
  the pattern indicated by the black dots at the various numbered
  regions. The crosses represent the beam position after the first
  reflection at the mirror-pair in region 10. (b) Similar mirror-pair
  scheme as in Fig.~\ref{fig:4} for 20 passes.  (c) The merging of the
  multi-passes scheme given in previous panels produces a beam
  propagation pattern of 100 passes.
  (d) 3D representation of the 3 mirror-pairs needed to realize the 100 passes
  scheme of panel (c).  The red circles represent the position of the
  disk,  the blue, green, magenta and cyan contours represent the
  first, second, third and forth mirror-pairs, respectively.}
\end{figure}

By merging the two concepts presented in Fig.~\ref{fig:3} and
Fig.~\ref{fig:4}, we can realize a novel multi-pass scheme with an increased
number of passes while at the same time making efficient use of the
surface of the parabolic mirror.
Before proceeding to develop in a detailed way the ideas behind the
novel configurations, the merging processes
yielding the simple layout of Fig.~\ref{fig:5} (d) is quickly
introduced.
The starting point is the setup of two mirror-pairs (green and blue)
as in Fig.~\ref{fig:5} (a) very similar to the setup in
Fig.~\ref{fig:3}.
Adding two more mirror-pairs (magenta and cyan) shown in
Fig.~\ref{fig:5} (b) results in an arrangement of four mirror-pairs as
depicted in Fig.~\ref{fig:5} (c) which can be simplified to the
three-mirror-pair setup shown in Fig.~\ref{fig:5} (d).

In more details, this merging proceeds in the following way. 
The panel (a) shows the same two mirror-pairs-arrangement as in
Fig.~\ref{fig:3} but contrarily to that situation, here the
beam is not coupled in the center of region 1 but with a small
displacement relative to the center of the region 1.
Starting from this initial position, the pump beam travels in the pump
optics following the given numbering at the positions indicated by the
black dots, until it eventually reaches region 10.
At this position, a third mirror-pair is placed, corresponding to the
magenta mirror-pair of panel (b).
This third mirror-pair reflects the beam within region 10, from the
black dot to the black cross.
From there, the beam undergoes a propagation indicated with the
crosses back to region 1.
At position 1 a fourth mirror-pair corresponding to the cyan
mirror-pairs of the panel  (b) is introduced, completing in this way
the merging of the two concepts.
The orientation of this forth mirror-pair is such to reflect the beam
within region 1 at the position indicated with the empty square.
Similar as before, starting from this position a propagation till the
region 10 is followed.
By iterating this scheme, a large number of passes  on the disk
can be realized.

Panel (c) of Fig.~\ref{fig:5} displays the arrangement of all optical elements resulting from the merging of
the two multi-pass schemes of panels (a) and (b).
The pump beam propagation follows the indicated numbering yielding
$N=2\times n\times m=2\times (180/\phi_1) \times (180/\phi_2)$ passes, where
$\phi_1$ and $\phi_2$ are the angular tilts in degrees of the second
and forth mirror-pairs, respectively.
Also in this case $N$ accounts for back-and-forth propagation in the
multi-pass segment.
The needed back-reflection at position 50 in panel (c) is realized by
the magenta mirror-pair.
The numbers $n$ and $m$, which represent the number of reflections at
the disk for the two separate multi-pass concepts, are used to classify
the various configurations resulting from the merging process.

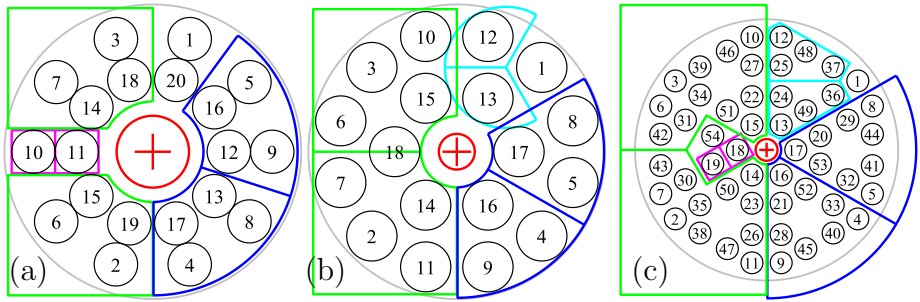
\begin{figure}[h]
\centering
    \setlength{\unitlength}{1.0mm}
    \begin{picture}(120,38)(0,0)
     %\put(0,0){\framebox(120,38)}
     \put(-1,-1){\includegraphics[width=0.33\textwidth]{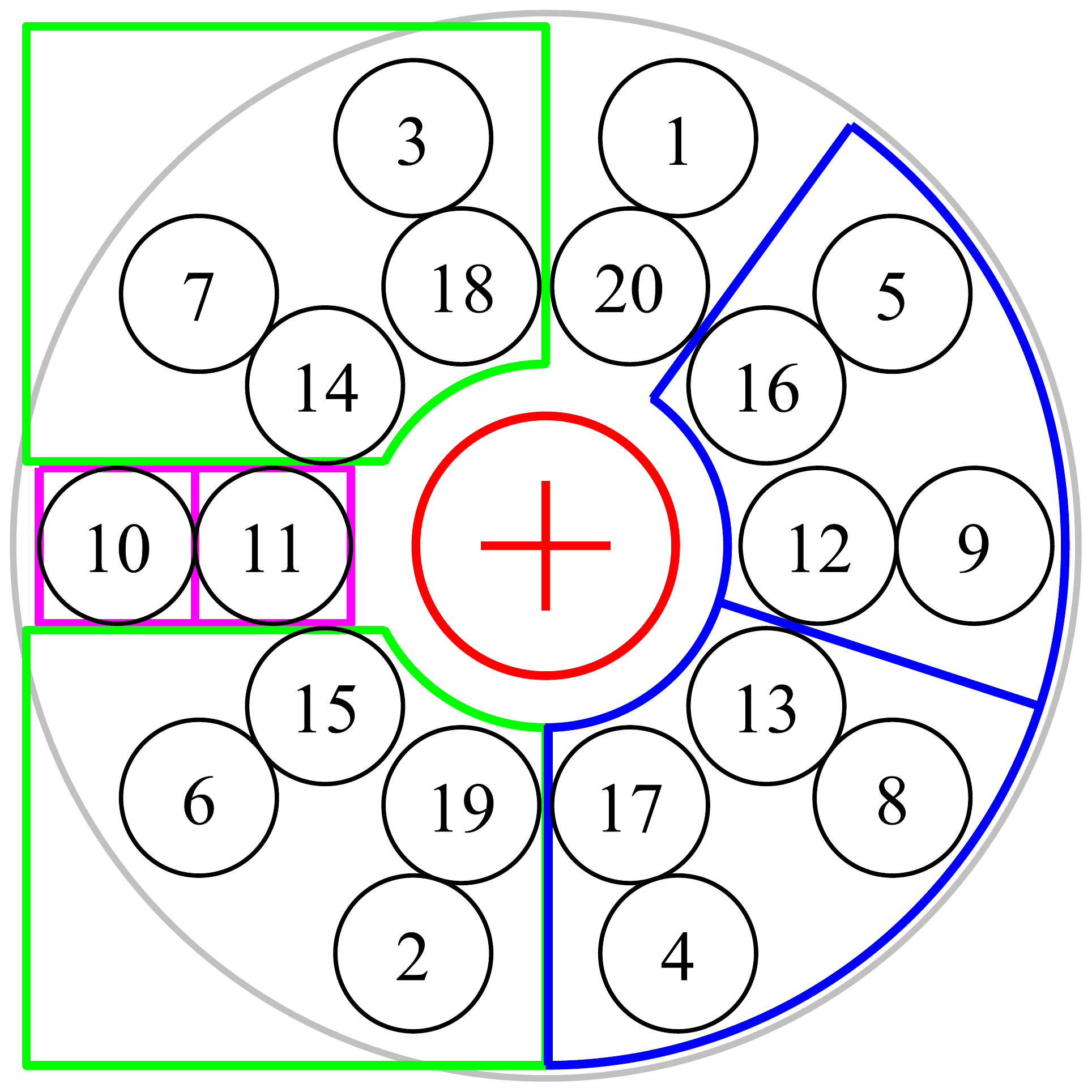}}
     \put(39,-1){\includegraphics[width=0.33\textwidth]{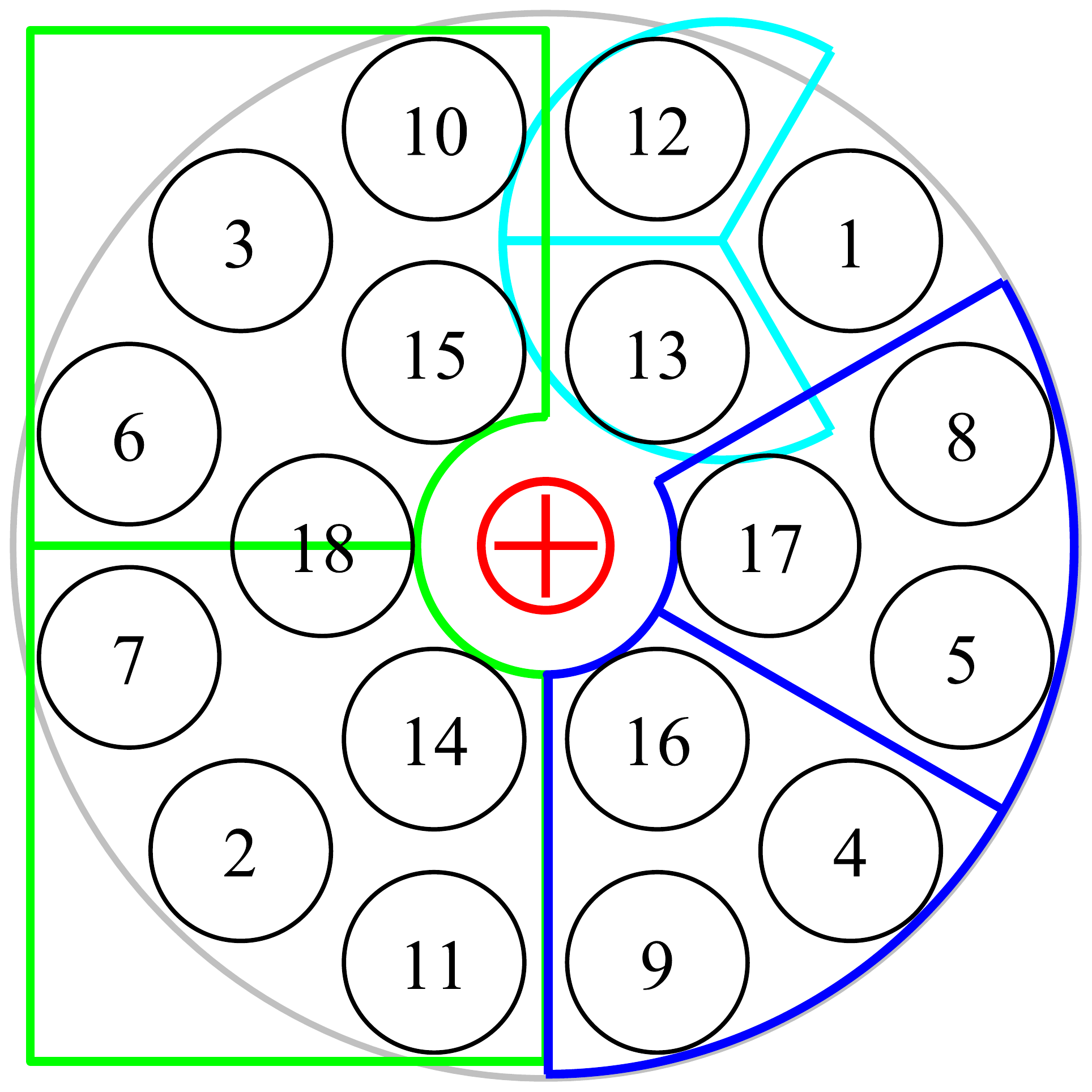}}
     \put(80,-1){\includegraphics[width=0.33\textwidth]{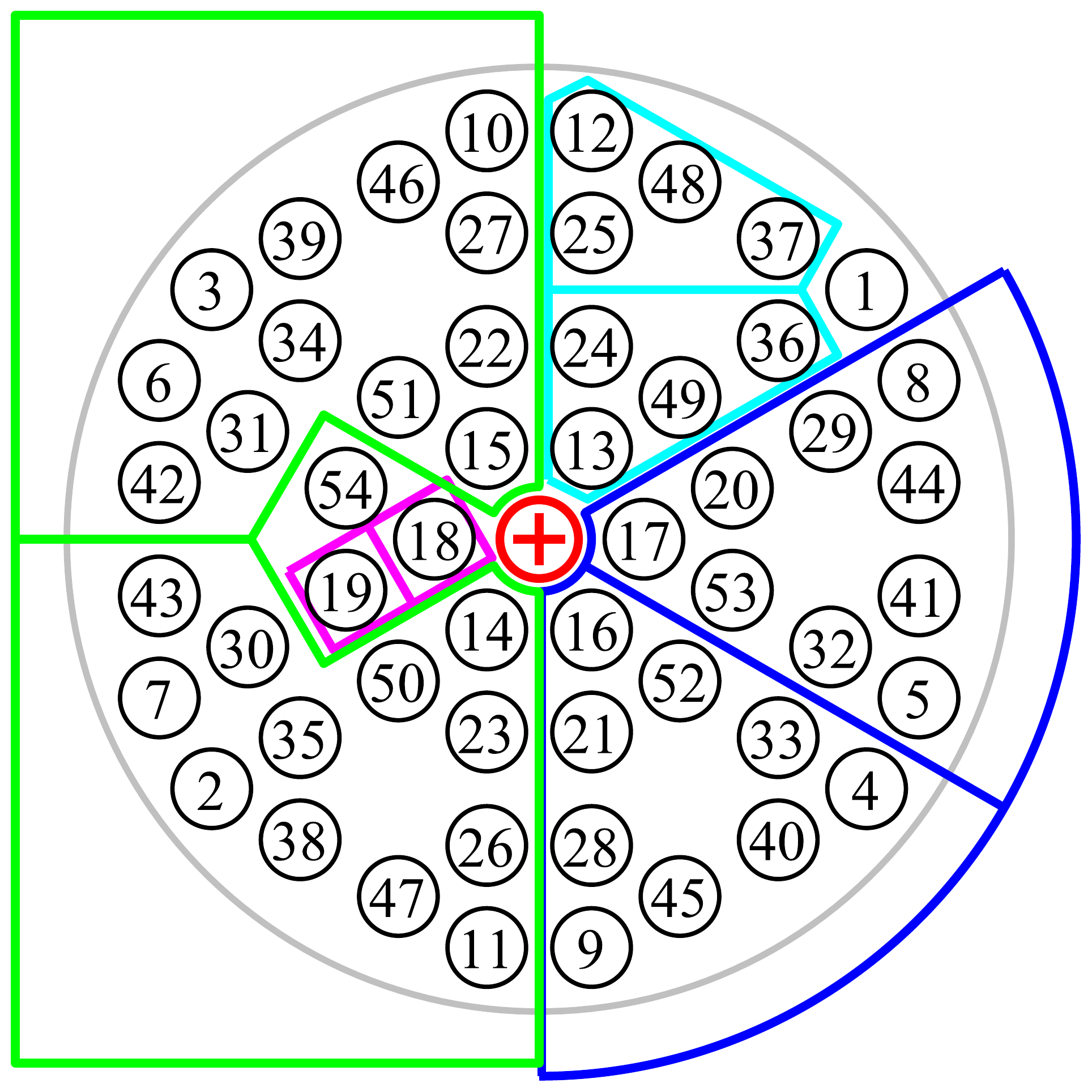}}
    \put(0,2){\large (a)}
    \put(39,2){\large (b)}
    \put(82,2){\large (c)}
    %\put(70,5){\large (d)}
\end{picture}
\caption{\label{fig:6} \it (a) $10\times 2$ configuration yielding 40
  passes. The third mirror pair in this case has been rotated by
  $90^\circ$ compared to the merging process described above. (b)
  $6\times 3$ configuration yielding 36 passes as implemented by
  TRUMPF, and (c) $6\times 3\times 3$ configuration which results by
  iterating twice a merging process resulting in 108 passes.}
\end{figure}

In this case, the third (magenta) mirror-pair was introduced just
for didactic reasons: to better highlight the principle of the merging
process.
In fact, in this particular case its functionality can be realized by
the first mirror-pair (see panel (d)), but this does not apply
generally (see e.g. Fig.~\ref{fig:6} (a)).

In conclusion, by adding only a small mirror-pair in the region of the
in-coupled beam (see Fig.~\ref{fig:5} (d)), the number of pump passes
relative to a standard pump optics design can be increased by a factor
of $(180/\phi_2)$.
Moreover, the individual beam spots at the mirror-pair and at the
parabolic mirror are larger than in the standard configuration
assuming same parabolic mirror and number of passes, because of the
better usage of the parabolic mirror surface.
Hence, the demands for the pump source phase-space quality or pump
optics size are decreased relative to the standard configurations.

The merging described here can be interpreted as a generalization of
the concept exploited already by TRUMPF~\cite{Alexander2012_patent6}
and shown in Fig.~\ref{fig:6} (b).
In our scheme, the TRUMPF layout can be classified as a $n\times
m=6\times 3$ configuration leading to 36 passes in the disk.
In panels (a) a $10\times 2$ and in panel (c) a $6\times 3\times 3$
configuration is shown. 
The latter demonstrates that our merging processes can be iterated
several times.
Moreover our merging process can be applied for various initial
configurations.

\begin{figure}[bt]
\centering\includegraphics[width=\textwidth]{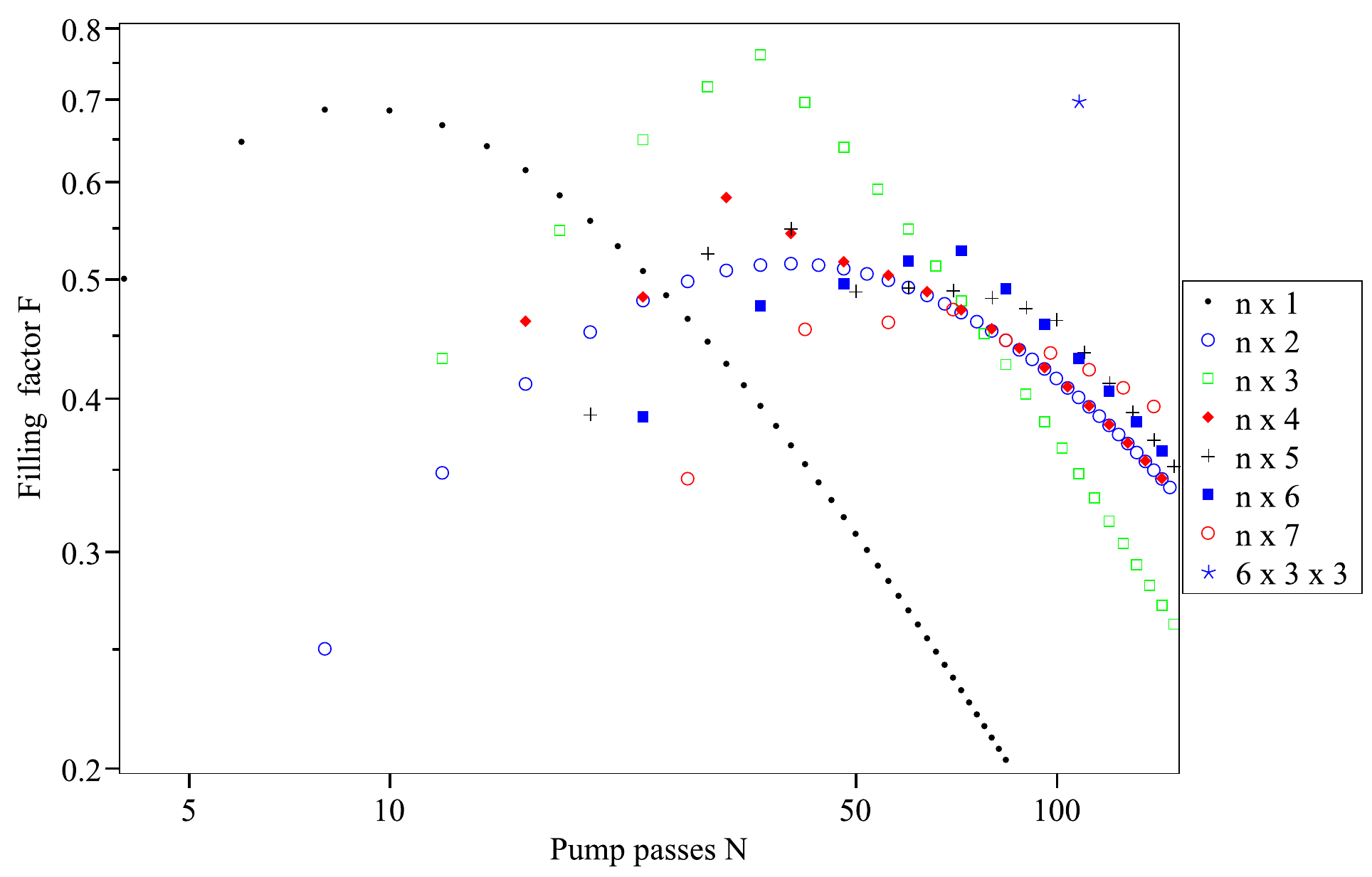}
\caption{\label{fig:7} \it ``Filling factor'' $F$ for various
  configurations versus number of passes $N$ classified using the
  $n\times m$ nomenclature and computed using the paraxial
  approximation. The $n\times 1$ configuration represents the standard
  layout of Fig.~\ref{fig:3}, the TRUMPF design corresponds to the maximum of the $n\times 3$ configuration and the
  star represents the $6\times 3\times3$ configuration of Fig.~\ref{fig:6}
(c). }
\end{figure}

A comparison of the various configurations is shown in
Fig.~\ref{fig:7}, where the ``filling factor'' $F$ is plotted against the
number of pump passes.
$F$ is defined as the ratio of the beam spots
area at the parabolic mirror over the parabolic mirror area:
\mbox{$F= Nd^2/(2D^2)$}, where $d$ is the individual beam spot
diameter in the far-field and $D$ the parabolic mirror diameter (see Fig.~\ref{fig:3} (c)).
$F$ has been computed using the paraxial approximation.  For the real
situation, the ``filling factor'' differences between the various
configurations are slightly larger due to the {\color{black} focal length  increase}
of the parabolic mirror with increasing distance from the axis.

For the standard design of Fig.~\ref{fig:3}, $D=2R+d$.
For large $N$, $D\approx 2R$ and $R \approx Nd/(2\pi)$ applies. Thus, in this limit  
the ``filling factor'' scales as $F\sim 1/N$.
For small $N$, these approximations do not apply and the ``filling
factor'' increases with $N$ as well visible from the plot.

Similar behavior versus $N$ is found for other configurations.
As a rule of thumb, with increasing $m$ the maximal ``filling factor''
is reached for larger values of $N$.
The overall maximal ``filling factor'' is obtained for the $N=36$
design, resulting from the $6\times 3$ configuration of TRUMPF
depicted in Fig.~\ref{fig:6} (b).
For $N=108$, the configuration $6\times 3\times3$ of Fig.~\ref{fig:6}
(c) shows by far the highest ``filling factor''.

Knowing the ``filling factor'' and the number of passes of a given
configuration, it is possible to determine the minimal diameter $D$ of
the parabolic mirror and the maximal beam parameters product $P$ of the
pump source:
\begin{eqnarray*}
D &>& \sqrt{\frac{8 N}{F}} \frac{Pf}{d_\mathrm{pump}}  \\
P &<& \sqrt{\frac{F}{8N}} \frac{D d_\mathrm{pump}}{f}
\end{eqnarray*}
where $d_\mathrm{pump}$ is the pump spot diameter at the disk and $f$
the focal length of the parabolic mirror.

The optimal scheme that is used in each particular application is a
trade-off between the number of passes needed, pump module size and
the quality and costs of the available pump source.

{\color{black} 
Another aspect that must be considered is the misalignment sensitivity
of the pump optics and the related required manufacturing precision.
Figure~\ref{fig:misalignement} shows misalignment plots for various
pump schemes.
%% %
\begin{figure}[h!]
\centering\includegraphics[width=\textwidth]{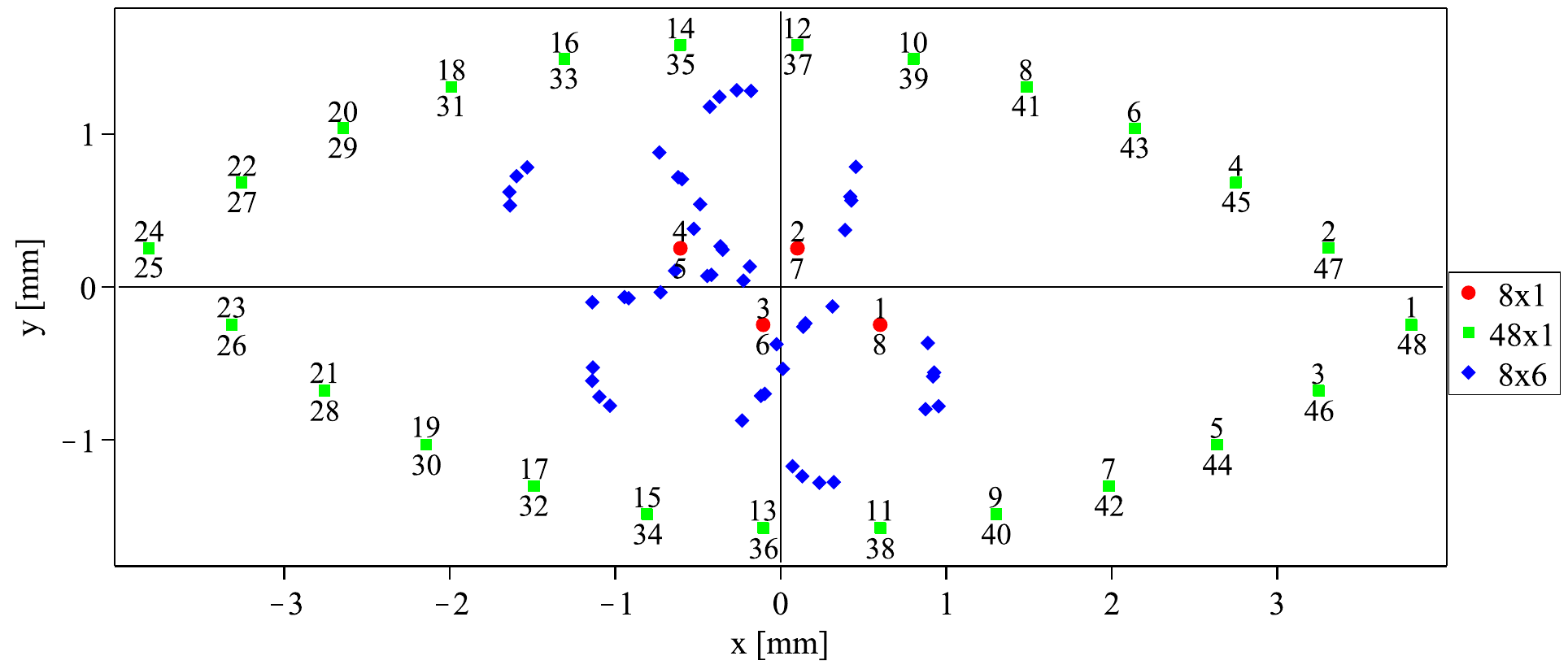}
\caption{\label{fig:misalignement} \it (Color online) Example of
  misalignment plot showing the deviation of the various pump beam
  passes at the thin-disk position for a departure of the angle
  between the two mirrors forming the first mirror-pair (given in
  green in all figures) by 2.5 mrad from 90 degrees. The parabolic
  mirror has a focal length of 200~mm.  The red points represent the
  deviation for the various passes in the standard $8\times 1$ design
  with $N=16$, the green squares for the standard $48\times 1$ design
  with $N=96$, and the blue diamonds for our $8\times 6$ configuration
  also with $N=96$.  The numbering which for clarity has been applied
  only to the standard configurations represents the pass number along
  the pump beam propagation. Similar plots are obtained for other
  misalignements.  }
\end{figure}
For an ideal alignment, each pump spot perfectly overlaps in the
center of the disk.
However practically, deviations from the ideal situation occur: for
example the  mirrors forming the mirror-pairs may not be exactly
orthogonal to each other, or the mirror-pair intersects may  not be
perfectly orthogonal to the symmetry axis, etc.
These misalignments give rise to deviations of the various pump spots
from the center of the disk.
Due to the rotation of the pump beam, compensation occurs while
propagating in the standard pump optics, so that the deviations of the
pump spots from the ideal position first increase and then decrease as
indicated by the numbering in Fig.~\ref{fig:misalignement}.
As the average deviation grows with the number of passes $N$, well
visible by comparing the red circles with the green squares,
misalignment effects becomes more severe for pump optics with large
number of passes.
However, in our design the large number of passes is reached starting
from a standard design with small number of passes.
Thus the deviations caused by misalignments are mitigated in our
schemes compared with the standard design having the same number of
passes, so that the presently used manufacturing precision is
sufficient.
}

\section{A particular example based on a  $6\times 6$ configuration  with triangular input aperture}
\label{sec:hexagonal}

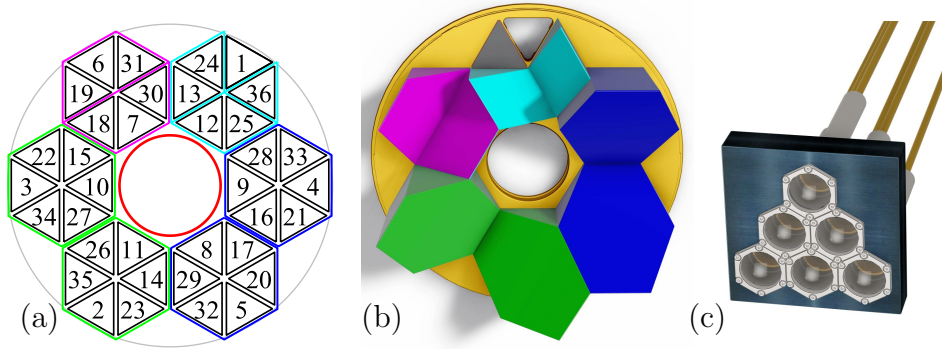
\begin{figure}[htbp]
%\centering\includegraphics[width=\textwidth]{Fig8_v1.jpg}
\centering
    \setlength{\unitlength}{1.0mm}
    \begin{picture}(120,45)(0,0)
     %\put(0,0){\framebox(120,45)}
     \put(-2,0){\includegraphics[width=0.36\textwidth]{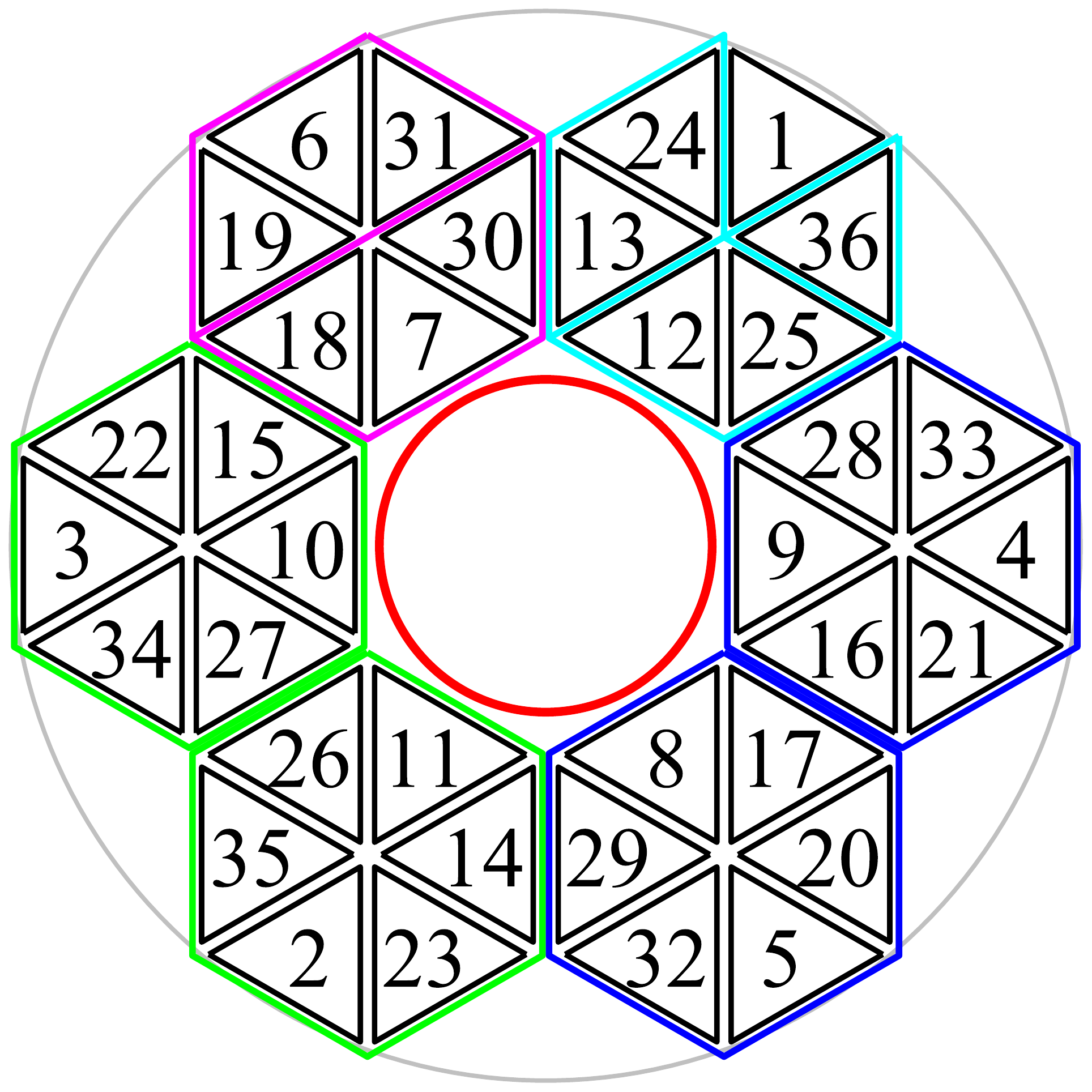}}
     \put(45,0){\includegraphics[width=0.36\textwidth]{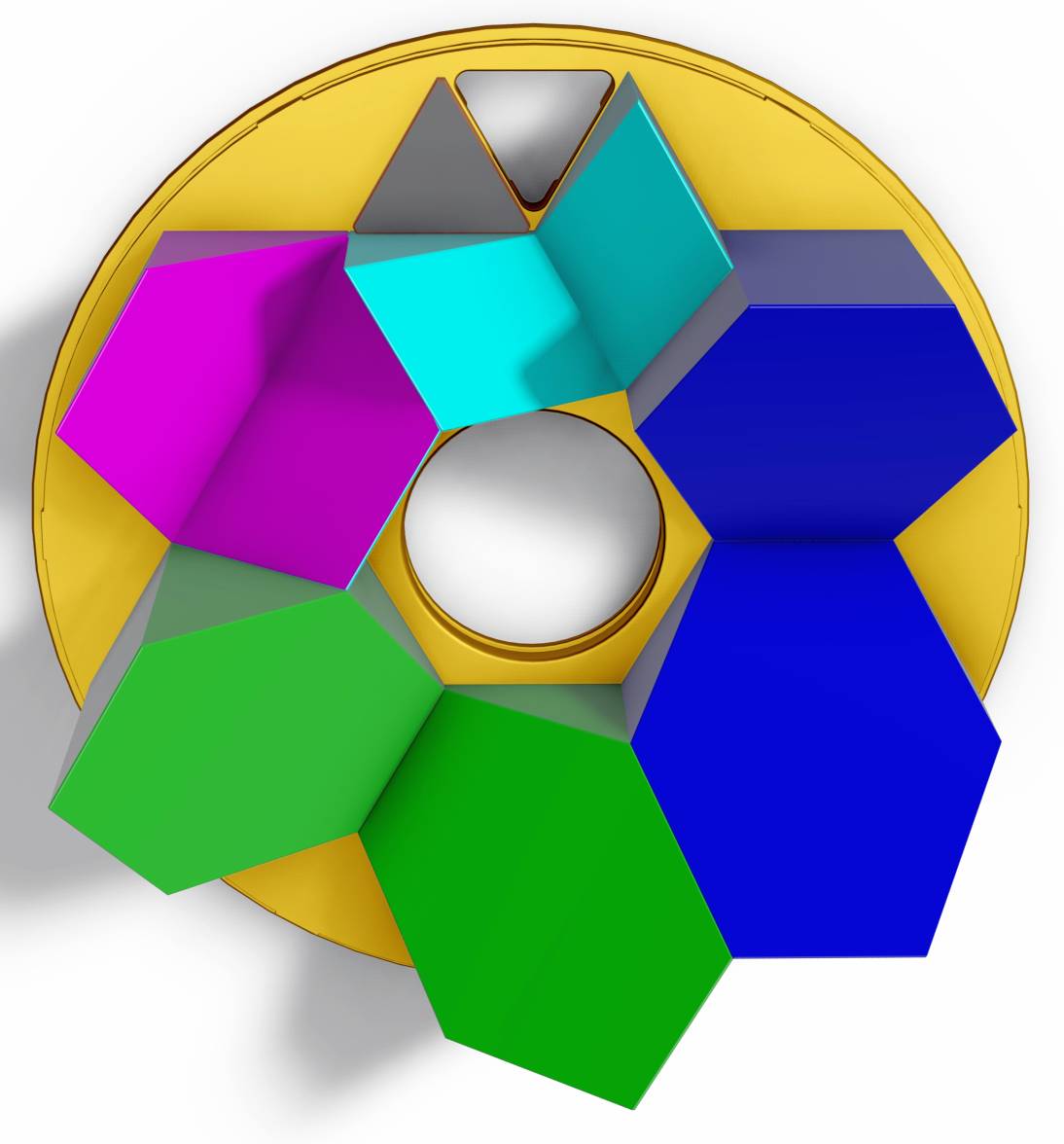}}
     \put(91,3){\includegraphics[width=0.25\textwidth]{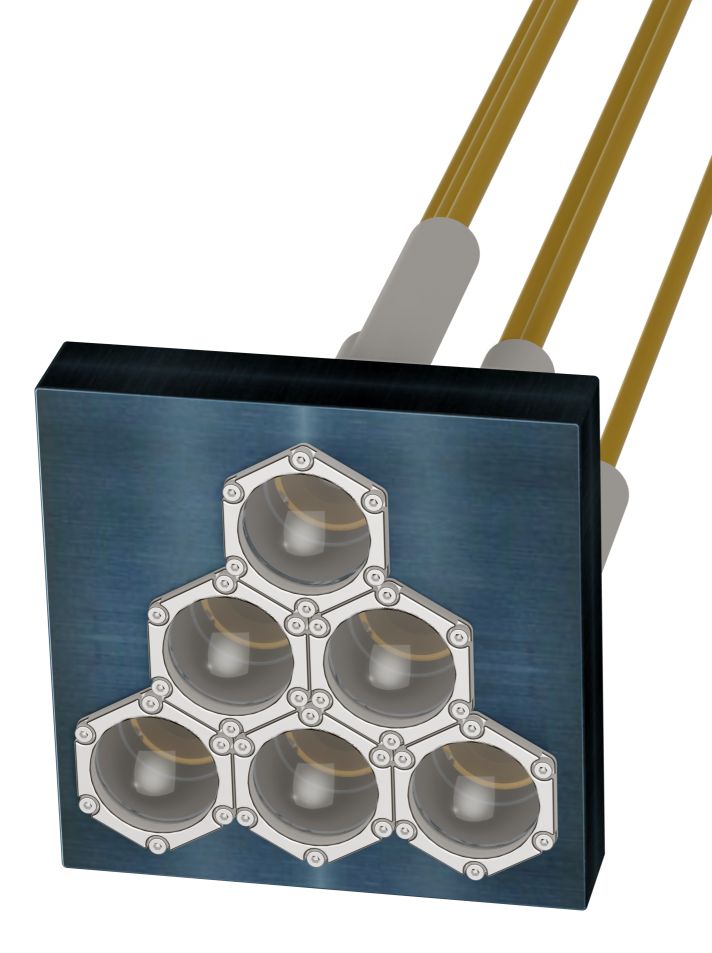}}
    \put(-0,3){\large (a)}
    \put(45,3){\large (b)}
    \put(88,3){\large (c)}
    %\put(70,5){\large (d)}
\end{picture}
\caption{\label{fig:8} \it (a) $6\times 6$ configuration yielding 72
  passes based on hexagonal mirror-pairs. (b) 3D drawings of the
  mirror-pairs.  This hexagonal shaped mirror-pairs are particularly
  suited for a triangular illumination which can be formed by merging
  several collimated beams arranged on an hexagonal lattice as shown in
  (c) and imaging them after appropriate magnification at the input
  beam triangular aperture (at position 1).}
\end{figure}

A realization of a 72 passes pump optics resulting from a $6\times 6$
configuration is presented in Fig.~\ref{fig:8}.
Two large hexagonal mirror-pairs, two small mirror-pairs and a flat
end-mirror are used in this configuration.
The hexagonal mirrors, placed as shown in Fig.~\ref{fig:8} (a) and
(b), naturally give rise to the central aperture necessary for the
laser beam without the need of intensive manufacturing related with
``internal'' cutaways as e.g., necessary for the prisms of
Fig.~\ref{fig:5} (d).
The beam losses occurring at position 6 of Fig.~\ref{fig:3} (b) or at
position 50 of Fig.~\ref{fig:5} (c) at the mirror-pair intersects are
not present in this configuration because at position 36 we are using
simply a flat mirror as back-reflector~\footnote{ {\color{black} When
    high power is requested, frequently the commercially available
    pump optics based on Fig.~\ref{fig:3} (b) design, at position 6
    (or equivalent positions) are equipped with a back-reflector to
    avoid losses which would be caused by the mirror-pair intersect.}
}, qualifying this scheme for improved efficiency and power scaling.

The triangular input aperture, which is successively imaged on the
various hexagonal mirrors seems strange for pump optics design since the
profile of commonly used homogenized pump radiation is usually
rotationally symmetric.
However, when high pump power is required, several collimated parallel
beams could be combined to illuminate the triangular input aperture.
The most homogeneous way to illuminate the triangular input aperture
is achieved by combining 3, 6, 10, 15...  pump diode homogenizers
outputs, placed on an hexagonal grid as shown in Fig.~\ref{fig:8}
(c).
Even though at the input-beam aperture (far-field) the various
parallel beams are located at different positions, the parabolic
mirror merges them into a single round spot at the disk position.
Large pump power density at the disk can be reached in this way, with
pump diameter given by the individual pump source outputs
characteristics (beam parameters product $P$).

\section{Conclusions}

A general scheme for multi-pass 4f pump optics suited for thin-disk
lasers has been presented here, which can accommodate for large number
of passes while keeping the requirements for the pump source beam
quality moderate.
A particular realization of this general scheme can be simply achieved
by inserting a small additional mirror pair at the input beam position
of a standard pump design.
In such a way, the same pump optics can be operated, on one hand,
without this new additional mirror-pair, for example, for high energy
pulsed pumping and thick disks.  On the other hand, by inserting a
mirror-pair at the in-coupling position tilted by 90, 60 and 30
degrees, the number of passes can be increased by factors of  2, 3 and 6,
respectively, for the case that larger numbers of passes are needed.
Thus, it is straightforward to adapt the number of passes to the given
pump source quality and active medium properties.
We also presented some simple formulae which can be used to estimate the
maximal beam parameters product a pump source must have, or the minimal
parabolic mirror diameter given a certain configuration and number
of passes.

The increased number of passes achievable by  using the 
schemes presented here, which is particularly important for active media having
small single pass absorption (due to lower disk thickness, lower
doping, lower active medium absorption cross section, and smaller
absorption bandwidth), yields an increased efficiency, enabling power
scaling, lower lasing threshold and the usage of novel materials.

This work is supported by the SNF\_200021L-138175, the
SNF\_200020\_159755  and the ERC StG. \#279765.

% \bibliographystyle{osajnl}
% \bibliography{/home/aldo/Dropbox/Work/JabRef/database}

%%%%%%%%%%%%%%%%%%%%%%% References %%%%%%%%%%%%%%%%%%%%%%%%%

 %%%%%%%%%%%%%%%%%%%%%%%%%%  body  %%%%%%%%%%%%%%%%%%%%%%%%%%

\end{document}